\documentclass{article}
\usepackage{spconf,amsmath,graphicx}
\usepackage{cite,amssymb,amsfonts,algorithmic,textcomp,xcolor,booktabs,multirow}
\DeclareRobustCommand{\ff}{\ensuremath{\text{f--f}}}
\DeclareRobustCommand{\savg}{\ensuremath{\text{s--avg}}}

\title{DAME: Duration-Aware Matryoshka Embedding for Duration-Robust Speaker Verification}
\name{Youngmoon Jung*, Joon-Young Yang*, Ju-ho Kim, Jaeyoung Roh, Chang Woo Han, Hoon-Young Cho\thanks{*Equal contribution.}}
\address{AI Solution Team, Samsung Research, Seoul, South Korea}
\begin{document}
\ninept
\maketitle
\begin{abstract}
Short-utterance speaker verification remains challenging due to limited speaker-discriminative cues in short speech segments.
While existing methods focus on enhancing speaker encoders, the embedding learning strategy still forces a single fixed-dimensional representation reused for utterances of any length, leaving capacity misaligned with the information available at different durations.
We propose Duration-Aware Matryoshka Embedding (DAME), a model-agnostic framework that builds a nested hierarchy of sub-embeddings aligned to utterance durations: lower-dimensional representations capture compact speaker traits from short utterances, while higher dimensions encode richer details from longer speech.
DAME supports both training from scratch and fine-tuning, and serves as a direct alternative to conventional large-margin fine-tuning, consistently improving performance across durations.
On the VoxCeleb1-O/E/H and VOiCES evaluation sets, DAME consistently reduces the equal error rate on 1-s and other short-duration trials, while maintaining full-length performance with no additional inference cost.
These gains generalize across various speaker encoder architectures under both general training and fine-tuning setups.

\end{abstract}
\section{Introduction}
\label{sec:intro}

Speaker verification (SV) aims to validate a claimed identity based on the speaker's voice.
Although recent advances in speaker encoders and training strategies have substantially improved performance on full-length utterances \cite{nagrani2017voxceleb, kye2020meta, jung2020fpm,mun2022msska,chen2023res2net,zhang2023adaptive,thienpondt2023ecapa2,kim2024faexunet,chen2024eres2netv2,wang2024mvec,park2025dsmegs}, short speech segments still pose a major challenge, due to limited phonetic and acoustic variability.
To address this, prior studies have proposed robust modeling strategies under duration constraints.
Among them, multi-scale feature aggregation (MFA) combines outputs from different encoder layers to build more generalizable frame-level representations \cite{jung2020fpm,mun2022msska,chen2023res2net,kim2024faexunet,chen2024eres2netv2}.
While effective, MFA strictly operates at the feature level and still trains a fixed-dimensional embedding, which may not adapt to the varying amounts of speaker information available from utterances of different lengths.
Notably, prior studies \cite{Kanagasundaram2019, Zheng2020} have shown that lower-dimensional embeddings are effective for short-utterance SV, while higher-dimensional embeddings better capture long-duration speaker traits, implying the inefficiency of fixed-size embedding learning against variable-length inputs.

Motivated by these observations, we introduce Duration-Aware Matryoshka Embedding (DAME), a model-agnostic framework that shifts the focus from feature aggregation to embedding extraction. 
To address the short-duration SV problem, DAME constructs a nested hierarchy of speaker embeddings aligned to utterance durations.
During training, input utterances are truncated to different lengths and then processed through a shared speaker encoder.
Subsequently, the sub-dimensions of the embeddings are trained such that lower- and higher-dimensional sub-embeddings are supervised to discriminate speakers using short- and longer-duration utterances, respectively.
By aligning the embedding sub-dimensions in proportion to the information available in the input utterances, this duration-aware supervision strategy encourages lower dimensions to capture compact speaker traits in short utterances while higher dimensions encode richer speaker information in longer utterances.

Because the proposed DAME framework is model-agnostic, it can also be applied to fine-tune pretrained speaker encoders with various architectures.
Conventional large-margin fine-tuning (LMFT) \cite{thienpondt2021idlab} is known to bias optimization toward long utterances and often degrades SV performance on short utterances \cite{zhang2023adaptive, thienpondt2023ecapa2}. 
Recent remedies like adaptive LMFT (ALMFT) \cite{zhang2023adaptive} and the ECAPA2 \cite{thienpondt2023ecapa2} training strategy mitigate this issue through margin scheduling and variable-duration training. 
Building on these advances, DAME offers a more direct solution at the embedding level. 
It replaces the single fine-tuning objective with multiple duration-supervised prefix losses on the nested embeddings, coupled with a duration-aligned, prefix-specific margin scheduling and variable-length mini-batches.
In our experiments, consistent gains are observed across three popular speaker encoders, ResNet34 \cite{chung2020metric}, ECAPA-TDNN \cite{Desplanques2020ecapatdnn}, and ERes2NetV2 \cite{chen2024eres2netv2}, demonstrating the generality and effectiveness of our method.

\section{Related Work}
\label{sec:related}

\noindent\textbf{Matryoshka Representation Learning.}
Matryoshka Representation Learning (MRL) \cite{kusupati2022mrl} was initially proposed in NLP to produce a high-dimensional embedding whose leading prefixes (e.g., first 32, 64, and 128 dimensions) could serve as sub-embeddings, enabling dynamic adjustment of embedding size without extra inference cost. 
In SV, M-Vec \cite{wang2024mvec} preserved SV performance with compact prefixes, and DSM-GS \cite{park2025dsmegs} improved training via dimension-aware margins and gradient scaling, yet these works mainly target efficiency rather than duration robustness.
We instead repurpose MRL for \emph{duration-aware} learning.
Motivated by evidence that short utterances prefer compact embeddings while long utterances benefit from higher dimensions \cite{Kanagasundaram2019, Zheng2020}, DAME aligns each duration scale to a matched-capacity sub-embedding.

\smallskip
\noindent\textbf{Multi-scale Feature Aggregation.}
MFA has been widely adopted to improve short-utterance SV robustness at the feature level. 
FPM \cite{jung2020fpm}, ERes2NetV2 \cite{chen2024eres2netv2}, and FA-ExU-Net \cite{kim2024faexunet} enhance encoder designs to integrate features across different temporal scales. 
However, these methods still train a fixed-dimensional embedding regardless of the input utterance length.
In contrast, the proposed DAME approach is independent of the speaker encoder architecture and instead restructures the embedding such that the sub-dimensions are aligned to the amount of speaker information available from the input utterance.
This makes the proposed DAME framework orthogonal and complementary to the conventional MFA methods.

\smallskip
\noindent\textbf{Large-Margin Fine-Tuning.}
Conventionally, the LMFT \cite{thienpondt2021idlab} strategy has been widely used to boost SV performance, but recent studies such as ALMFT \cite{zhang2023adaptive} and ECAPA2 \cite{thienpondt2023ecapa2} reported that LMFT can degrade short-utterance SV performance.
This was because the LMFT process focused on the optimization of model parameters toward long utterances with a larger margin.
To address this problem, a variable-duration LMFT method was presented in \cite{zhang2023adaptive,thienpondt2023ecapa2}, which composed training mini-batches with utterances of variable durations from shorter to longer ones.
Specifically, a duration-based ALMFT (D-ALMFT) strategy was proposed in \cite{zhang2023adaptive}, which adaptively introduced smaller and larger margins for shorter and longer utterances, respectively.

\begin{figure}[t]
  \centering
  \includegraphics[width=0.7\linewidth]{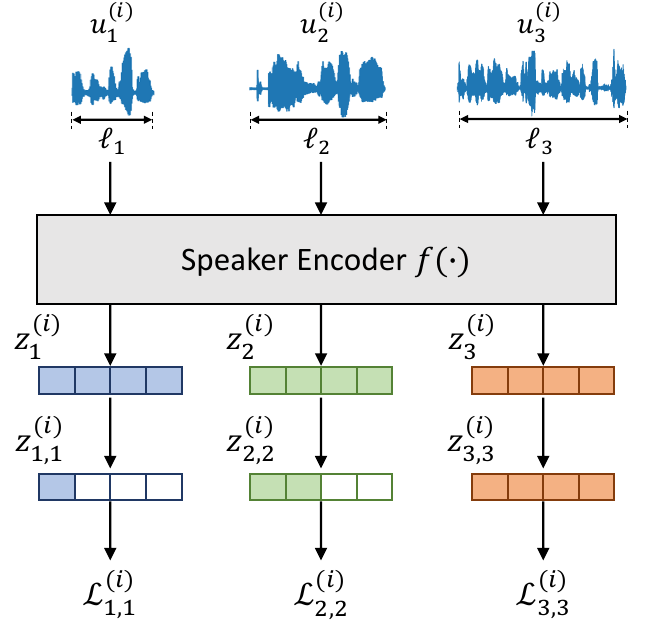}
  \vspace{-0.15cm}
  \caption{DAME overview (HW, $J{=}K{=}3$): one-to-one duration–prefix alignment; each chunk supervises its matched sub-embedding via a large-margin loss.}
  \label{fig:fig1}
  \vspace{-0.25cm}
\end{figure}

\section{Proposed method}
\label{sec:method}

\subsection{Overview and Notation}

During training, let $y^{(i)}$ denote the speaker label for the $i$-th training instance, and let $\mathcal{T}=\{\ell_1,\ldots,\ell_J\}$ be the set of target chunk durations with $\ell_1<\cdots<\ell_J$. 
For each instance $i$, we construct $J$ speech chunks $\{u^{(i)}_j\}_{j=1}^{J}$ by truncating or cropping different utterances of the same speaker to durations $\ell_j\in\mathcal{T}$. 
Each chunk is passed through a shared encoder $f(\cdot)$ to produce a full speaker embedding $z^{(i)}_j=f(u^{(i)}_j)\in\mathbb{R}^{D}$, where $D$ is the full embedding dimension. 
Given a set of prefix (nesting) dimensions $\mathcal{D}=\{ d_1, \ldots, d_K \}$ with $d_1<\cdots<d_K$ and $d_K=D$, we define the prefix (sub-)embeddings $z^{(i)}_{j,k}\in\mathbb{R}^{d_k}$ as the leading $d_k$ components of $z^{(i)}_{j}$.

\subsection{DAME Loss}
\label{subsec:dame_loss}

In the proposed DAME framework, each prefix embedding $z^{(i)}_{j,k}$ is supervised by a large-margin classifier (SphereFace2 \cite{Bing-ICASSP}). The per-prefix loss is defined as
\begin{equation}\label{eq:single_prefix_loss}
\mathcal L^{(i)}_{j,k}
= \mathcal L_{\mathrm{cls}}\!\big(z^{(i)}_{j,k},\, y^{(i)};\, s, m_{d_k}\big).
\end{equation}
Here, $s=30$ is the fixed logit scale, and $m_{d_k}$ is the angular margin for the prefix dimension $d_k$.
The multi-prefix loss aggregates the per-prefix losses using duration--dimension alignment weights $c_{j,k}$:
\begin{align}\label{eq:multi_prefix_loss}
\mathcal L^{(i)}_{j,\mathcal{D}} \;=\; \sum_{k=1}^{K} c_{j,k}\, \mathcal L^{(i)}_{j,k}.
\end{align}
Finally, the DAME loss aggregates multi-prefix losses for all cropped chunks of all utterances in a mini-batch:
\begin{eqnarray}\label{eq:dame_loss}
\mathcal L_{\mathcal{T},\mathcal{D}}
= \frac{1}{I} \sum_{i=1}^{I} \left(
  \alpha \, \mathcal L^{(i)}_{J,\mathcal{D}}
  + \frac{1-\alpha}{J-1} \sum_{j=1}^{J-1} \mathcal L^{(i)}_{j,\mathcal{D}}
\right),
\end{eqnarray}
where $I$ is the mini-batch size, and $\alpha$ is a weighting factor between the longest chunk ($j=J$) and the remaining chunks ($j{=}1,\ldots,J{-}1$).

As explained briefly in Section \ref{sec:intro}, the DAME approach aims to align lower- and higher-dimensional sub-embeddings with shorter and longer chunks, respectively.
To achieve this, the duration--dimension alignment weights, $c_{j,k}$, in Eq.~(\ref{eq:multi_prefix_loss}) are defined as follows:
\begin{equation}\label{eq:alignment_weights}
c_{j,k} =
\begin{cases}
1, & b_{j-1} < k \le b_{j},\\
\gamma_k, & \text{otherwise}.
\end{cases}
,\;\;b_j=\Big\lfloor j \cdot \frac{K}{J} \Big\rfloor,\;\; b_0=0,
\end{equation}
where $j\in\{1,\dots,J\}$ and $k\in\{1,\dots,K\}$.
This definition yields a disjoint partition of $\{1,\ldots,K\}$ with no overlaps or gaps.
Using Eq.~\eqref{eq:alignment_weights}, $c_{j,k}{=}1$ for prefixes in the $j$-th band $(b_{j-1},\,b_j]$ (i.e., those near the scaled diagonal $k\!\approx\!\tfrac{jK}{J}$), and $c_{j,k}{=}\gamma_k$ otherwise.

Two weighting schemes are considered in Eq.~\eqref{eq:alignment_weights}.
In the hard weighting (HW) case, we set $J{=}K$ and $\gamma_k{=}0$, enforcing a strict one-to-one correspondence between the cropped durations $\{\ell_1,\ldots,\ell_J\}$ and the prefix dimensions $\{d_1,\ldots,d_K\}$, such that each $\ell_j$ supervises only $d_j$.
In the soft weighting (SW) case, we set $J{<}K$ and $\gamma_k{=}2^{-(K-k+1)}$, so that $0{<}\gamma_k{\le}\frac{1}{2}$ and $\gamma_k$ increases monotonically with $k$.
This setup keeps the near-diagonal band at $1$, while softly down-weighting entries outside this band as the prefix dimension $d_k$ decreases, thus promoting gradual cross-dimension transfer.
With these weighting schemes, the proposed DAME strategy (Eqs.~(\ref{eq:single_prefix_loss})--(\ref{eq:alignment_weights})) promotes a hierarchical supervision between the predefined duration and dimension sets, thereby encouraging lower prefixes to learn compact cues from short utterances and higher prefixes to encode richer information from longer utterances.
Figure~\ref{fig:fig1} illustrates the HW case with $J{=}K{=}3$.

\begin{table*}[t]
\centering
\caption{DAME configuration for each encoder architecture under general training and fine-tuning setups.}
\vspace{+0.05cm}
\label{tab:dame-config}
\scriptsize
\renewcommand{\arraystretch}{1.1}
\begin{tabular}{c|c|c|c|c|c}
\toprule
\textbf{Weighting Scheme} & \textbf{Encoder Architecture} & \textbf{Nesting Dimension Set ($\mathcal{D}$)} & \textbf{Duration Set ($\mathcal{T}$)} & $\{m_{d_k}\}_{\text{init}}$ & $\{m_{d_k}\}_{\text{final}}$ \\
\midrule
\multirow{3}{*}{SW}
& ResNet34      & \{32, 64, 128, 256\}         & \{1.0, 2.0\} & \{0.0, 0.0, 0.0, 0.0\}          & \{0.0, 0.1, 0.2, 0.2\} \\
& ECAPA-TDNN    & \{24, 48, 96, 192\}         & \{1.0, 2.0\} & \{0.0, 0.0, 0.0, 0.0\}          & \{0.0, 0.0, 0.1, 0.2\} \\
& ERes2NetV2    & \{24, 48, 96, 192\}    & \{1.0, 2.0\} & \{0.0, 0.0, 0.0, 0.0\}     & \{0.0, 0.0, 0.1, 0.2\} \\
\midrule
\multirow{3}{*}{HW}
& ResNet34      & \{64, 128, 256\}         & \{1.0, 2.0, 6.0\} & \{0.0, 0.2, 0.5\}          & \{0.0, 0.2, 0.5\} \\
& ECAPA-TDNN    & \{48, 96, 192\}         & \{1.0, 2.0, 6.0\} & \{0.0, 0.2, 0.5\}          & \{0.0, 0.2, 0.5\} \\
& ERes2NetV2    & \{48, 96, 192\}    & \{1.0, 2.0, 3.0\} & \{0.0, 0.2, 0.4\}     & \{0.0, 0.2, 0.4\} \\
\bottomrule
\end{tabular}
\vspace{-0.25cm}
\end{table*}

\subsection{Training Regime}
\label{subsec:ft_dame}

The proposed DAME framework can be applied either to train a speaker encoder from scratch or to fine-tune a pretrained model.
We refer to these two schemes as DAME \emph{general training} (DAME-GT) and DAME \emph{fine-tuning} (DAME-FT), respectively.
For DAME-GT, following the original MRL formulation~\cite{kusupati2022mrl}, each prefix dimension is assigned a separate classification head (i.e., weight matrix) $W_k \in \mathbb{R}^{d_k \times C}$,
where the column size $C$ denotes the number of training speakers.
Because the weights are not shared across the prefix dimensions, the training allows each sub-embedding to contribute better to the objective assigned to it.
For DAME-FT, we adopt the efficient \emph{MRL-E} weight-tying strategy proposed in~\cite{kusupati2022mrl}. We first train a speaker encoder using the standard approach with a single classification head $W \in \mathbb{R}^{D \times C}$.
Then, the classification head of a pretrained speaker encoder is shared among the different prefix dimensions by simply taking the leading $d_k$ rows of the weight matrix.
This weight-tying scheme improves training efficiency as it avoids training new heads for the prefix dimensions from scratch.

\section{Experimental Setup}

\subsection{Model Training}\label{sec:model-config}
All experiments used the WeSpeaker toolkit~\cite{wespeaker} on a single NVIDIA H100 GPU. We evaluated three encoders in their original configurations: ResNet34 ($D{=}256$, 6.4M)~\cite{chung2020metric}, ECAPA-TDNN ($D{=}192$, 15.2M)~\cite{Desplanques2020ecapatdnn}, and ERes2NetV2 ($D{=}192$, 34.8M)~\cite{chen2024eres2netv2}. Baselines followed WeSpeaker recipes with 2\,s crops, SphereFace2 loss~\cite{Bing-ICASSP} with a final margin of 0.2, SGD with an initial learning rate (LR) of 0.01, and the Triangular2 scheduler~\cite{Smith-WACV}.

For the proposed DAME approach, we adopted the HW and SW schemes described in Sec.~\ref{subsec:dame_loss}. 
Table~\ref{tab:dame-config} lists the nesting dimensions $\mathcal{D}$, the duration sets $\mathcal{T}$, and the per-prefix angular margins for each prefix dimension $d_k$, reported as initial and final values $\{m_{d_k}\}_{\text{init}}$ and $\{m_{d_k}\}_{\text{final}}$. 
These hyperparameters were determined by the following principles: 
(i) for HW, we reuse the conventional LMFT-style durations and margins~\cite{zhang2023adaptive,thienpondt2023ecapa2}; (ii) for SW, we extend the standard 2-s crop with an additional 1-s crop and assign a smaller angular margin to its corresponding low-dimensional prefixes (e.g., $m{=}0.1$ for 1-s), reflecting that shorter segments contain less speaker evidence and thus benefit from a softer constraint. 
Under the alignment weights in Eq.~\eqref{eq:alignment_weights}, this margin design couples duration and prefix dimension through the loss weighting: although $m_{d_k}$ is \emph{dimension-specific} (per-prefix), the losses for a chunk of duration $\ell_j$ are weighted toward its matched prefixes, yielding \emph{duration-aligned} effective margins (smaller for shorter crops, larger for longer crops). 
The heuristic of assigning smaller margins to lower-dimensional prefixes is motivated by dimension-aware margin scheduling in prior work~\cite{park2025dsmegs}. 
In all cases, we choose $(\mathcal{D},\mathcal{T})$ to fit the GPU memory budget and keep the number of distinct settings minimal. 


For DAME-GT, models were trained from scratch with Eq.~\eqref{eq:dame_loss}; $\alpha$ was initially set to $1.0$ and then linearly decayed to $0.5$ by epoch 50, and margins for each prefix $d_k$ were warmed up exponentially from $0.0$ to $m_{d_k}$ during epochs 30--40.
For DAME-FT, pretrained baselines were fine-tuned for 30 epochs using the DAME loss in Eq.~\eqref{eq:dame_loss} with fixed $\alpha{=}0.5$ and fixed final margins $m_{d_k}$ (Table~\ref{tab:dame-config}); the LR decayed exponentially from $10^{-4}$ to $10^{-5}$.

All models were trained on VoxCeleb2-dev~\cite{chung2018voxceleb2} at 16\,kHz. We applied on-the-fly data augmentation using MUSAN~\cite{Snyder-arxiv} and simulated room impulse responses and extracted 80-dim log Mel-filterbanks (25\,ms window, 10\,ms shift) with utterance-level mean subtraction.
Mini-batches contained 128 speakers for ResNet34 and ECAPA-TDNN, and 64 for ERes2NetV2. 
For DAME training, the $J$ chunks ($j{=}1,\dots,J$) were drawn from different utterances of the same speaker.

\subsection{Evaluation Protocols}
Evaluation was conducted on four test subsets: VoxCeleb1-O, VoxCeleb1-E, VoxCeleb1-H \cite{nagrani2017voxceleb}, and VOiCES \cite{Nandwana2019voices} corpus.
We followed the protocol in \cite{jung2020fpm} and considered multiple trial conditions with varied enrollment--test durations: full--full (`f--f'), 5s--5s, 5s--3s, 5s--2s, and 5s--1s.
This setup was intended to reflect practical scenarios in which an utterance of sufficient duration is typically required for speaker enrollment, but a short phrase is used in test time.
All systems—including MRL and DAME variants—use the full embedding at test time ($d_K{=}D$); thus the inference procedure and cost are unchanged.
We report equal error rate (EER) and, due to space, the macro-average over \{5s--5s, 5s--3s, 5s--2s, 5s--1s\} as `s--avg'.

\section{Results}

\begin{table}[t]
\centering
\scriptsize
\setlength{\tabcolsep}{2pt}
\vspace{-0.2cm}
\caption{General training results on VoxCeleb1 (EER, \%).}
\label{tab:gen-compact-split}
\vspace{+0.05cm}
\resizebox{\columnwidth}{!}{
\begin{tabular}{l l ccc ccc ccc}
\toprule
\multirow{2}{*}{Encoder} & \multirow{2}{*}{Method} &
\multicolumn{3}{c}{VoxCeleb1-O} & \multicolumn{3}{c}{VoxCeleb1-E} & \multicolumn{3}{c}{VoxCeleb1-H} \\
\cmidrule(lr){3-5}\cmidrule(lr){6-8}\cmidrule(lr){9-11}
& & \ff & \savg & 5s--1s
  & \ff & \savg & 5s--1s
  & \ff & \savg & 5s--1s \\
\midrule
\multirow{5}{*}{ResNet34}
& Baseline      & 1.02 & 2.24 & 4.33  & 1.12 & 2.27 & 4.25  & \textbf{2.03} & 3.77 & 6.57 \\
& +VLT      & 1.02 & 1.89 & 3.35  & \textbf{1.11} & 1.98 & 3.41  & 2.11 & 3.48 & 5.62 \\
& MRL           & 0.99 & 2.32 & 4.39  & 1.14 & 2.29 & 4.26  & 2.11 & 3.87 & 6.65 \\
& {DAME-GT-SW} & \textbf{0.92} & \textbf{1.82} & \textbf{3.29}  & 1.16 & \textbf{1.93} & \textbf{3.40}  & 2.11 & \textbf{3.46} & \textbf{5.59} \\
& DAME-GT-HW      & 0.98 & 2.11 & 3.89  & 1.15 & 2.19 & 3.92  & 2.10 & 3.71 & 6.13 \\
\midrule
\multirow{5}{*}{ECAPA-TDNN}
& Baseline      & 1.08 & 2.54 & 4.84  & \textbf{1.13} & 2.44 & 4.68  & \textbf{2.00} & 3.93 & 6.95 \\
& +VLT      & \textbf{1.00} & 2.15 & 3.97  & 1.18 & 2.20 & 3.92  & 2.19 & 3.83 & 6.34 \\
& MRL           & 1.03 & 2.41 & 4.69  & 1.19 & 2.46 & 4.66  & 2.20 & 4.16 & 7.26 \\
& {DAME-GT-SW} & 1.08 & \textbf{2.06} & \textbf{3.66}  & 1.19 & \textbf{2.13} & \textbf{3.69}  & 2.16 & \textbf{3.80} & \textbf{6.19} \\
& DAME-GT-HW      & 1.09 & 2.29 & 4.21  & 1.15 & 2.27 & 4.18  & 2.08 & 3.88 & 6.72 \\
\midrule
\multirow{5}{*}{ERes2NetV2}
& Baseline      & 0.91 & 2.04 & 3.89  & 1.15 & 2.26 & 4.11  & 2.09 & 3.81 & 6.40 \\
& +VLT      & 0.93 & 1.91 & 3.38  & 1.11 & 1.98 & 3.35  & 2.09 & 3.48 & \textbf{5.53} \\
& MRL           & \textbf{0.89} & 1.97 & 3.77  & \textbf{1.00} & 2.01 & 3.70  & \textbf{1.87} & 3.45 & 5.91 \\
& {DAME-GT-SW} & 0.91 & \textbf{1.82} & \textbf{3.26}  & 1.06 & \textbf{1.94} & \textbf{3.32}  & 2.04 & \textbf{3.44} & 5.59 \\
& DAME-GT-HW      & 0.93 & 1.90 & 3.55  & 1.04 & 2.05 & 3.44  & 1.98 & 3.55 & 5.81 \\
\bottomrule
\end{tabular}}
\vspace{-0.35cm}
\end{table}

\begin{table}[t]
\centering
\scriptsize
\setlength{\tabcolsep}{2pt}
\caption{Fine-tuning results on VoxCeleb1 (EER, \%).}
\label{tab:ft-compact-split}
\vspace{+0.1cm}
\resizebox{\columnwidth}{!}{
\begin{tabular}{l l ccc ccc ccc}
\toprule
\multirow{2}{*}{Encoder} & \multirow{2}{*}{Method} &
\multicolumn{3}{c}{VoxCeleb1-O} & \multicolumn{3}{c}{VoxCeleb1-E} & \multicolumn{3}{c}{VoxCeleb1-H} \\
\cmidrule(lr){3-5}\cmidrule(lr){6-8}\cmidrule(lr){9-11}
& & \ff & \savg & 5s--1s
  & \ff & \savg & 5s--1s
  & \ff & \savg & 5s--1s \\
\midrule
\multirow{5}{*}{ResNet34}
& Pretrained     & 1.02 & 2.24 & 4.33  & 1.12 & 2.27 & 4.25  & 2.03 & 3.77 & 6.57 \\
& LMFT  & 0.95 & 2.35 & 4.59  & 1.01 & 2.33 & 4.45  & 2.01 & 4.08 & 7.18 \\
& D-ALMFT   & 1.00 & 2.04 & 4.10  & 1.18 & 2.15 & 4.06  & 1.85 & 3.70 & 6.50 \\
& {DAME-FT-SW}    & 1.47 & 2.35 & \textbf{3.96}  & 1.49 & 2.36 & \textbf{3.91}  & 2.71 & 4.25 & 6.89 \\
& {DAME-FT-HW}    & \textbf{0.92} & \textbf{2.00} & 4.19  & \textbf{0.97} & \textbf{2.11} & 4.03  & \textbf{1.79} & \textbf{3.65} & \textbf{6.45} \\
\midrule
\multirow{5}{*}{ECAPA-TDNN}
& Pretrained     & 1.08 & 2.54 & 4.84  & 1.13 & 2.44 & 4.68  & 2.00 & 3.93 & 6.95 \\
& LMFT  & \textbf{0.94} & 2.80 & 5.67  & \textbf{1.00} & 2.70 & 5.58  & \textbf{1.91} & 4.31 & 8.10 \\
& D-ALMFT   & 1.05 & 2.28 & \textbf{4.30}  & 1.18 & 2.31 & \textbf{4.17}  & 2.15 & {4.01} & 6.90 \\
& {DAME-FT-SW}    & 1.13 & 2.43 & 4.51  & 1.18 & 2.32 & 4.27 & 2.13 & \textbf{3.92} & \textbf{6.78} \\
& {DAME-FT-HW}    & 1.00 & \textbf{2.27} & 4.39  & 1.14 & \textbf{2.30} & 4.22  & 2.10 & 4.06 & 6.80 \\
\midrule
\multirow{5}{*}{ERes2NetV2}
& Pretrained     & 0.91 & 2.04 & 3.89  & 1.15 & 2.26 & 4.11  & 2.09 & 3.81 & 6.40 \\
& LMFT  & 0.84 & 2.18 & 4.18  & \textbf{1.03} & 2.25 & 4.12  & 2.04 & 3.95 & 6.79 \\
& D-ALMFT   & 0.90 & 2.08 & 3.91  & 1.14 & 2.16 & 3.89  & 2.02 & \textbf{3.56} & 6.10 \\
& {DAME-FT-SW}    & 1.33 & 2.34 & 4.14  & 1.46 & 2.36 & 4.12  & 2.56 & 4.17 & 7.20 \\
& {DAME-FT-HW}    & \textbf{0.83} & \textbf{1.97} & \textbf{3.78}  & 1.06 & \textbf{2.09} & \textbf{3.77}  & \textbf{1.98} & 3.61 & \textbf{6.07} \\
\bottomrule
\end{tabular}}
\vspace{-0.3cm}
\end{table}

\subsection{Results on VoxCeleb1}

Table~\ref{tab:gen-compact-split} compares DAME-GT with baselines on the VoxCeleb1 test sets, including a variable-length training (VLT) baseline (identical to the baseline but trained with $\mathcal{T}{=}\{1.0,2.0\}$) and a plain MRL variant~\cite{wang2024mvec}. 
While \ff\ was broadly comparable across methods, both VLT and DAME-GT significantly outperformed the baseline and MRL on \savg\ and 5s--1s, attributable to short-duration supervision.
Although VLT and DAME-GT-SW were competitive, the latter exhibited 4.2\%--7.8\% lower EERs relative to the former on short-duration trials (s-avg and 5s--1s, respectively) with the ECAPA-TDNN encoder on VoxCeleb1-O.
Between SW and HW schemes, the DAME-GT-SW consistently outperformed the DAME-GT-HW on both s--avg and 5s--1s conditions.

Table~\ref{tab:ft-compact-split} summarizes fine-tuning results. 
We compared LMFT~\cite{thienpondt2021idlab} and D-ALMFT~\cite{zhang2023adaptive} as conventional fine-tuning baselines: LMFT used 6-s utterances with a fixed margin of 0.5, while D-ALMFT replaced the DAME-FT-HW nesting dimension set with $\mathcal{D}{=}\{D,D,D\}$. 
LMFT slightly improved the baseline system on the f--f condition but exhibited overall EER increases on both s--avg and 5s--1s conditions.
In contrast, both D-ALMFT and DAME-FT-HW generally improved the baseline on the overall test conditions, with the latter consistently outperforming the former with ResNet34 and ERes2NetV2 architectures.
DAME-FT-SW occasionally achieved the lowest EER on the 5s--1s condition, but tended to hurt the f--f performance.

\begin{figure}[t]
  \centering
  \includegraphics[width=0.9\columnwidth]{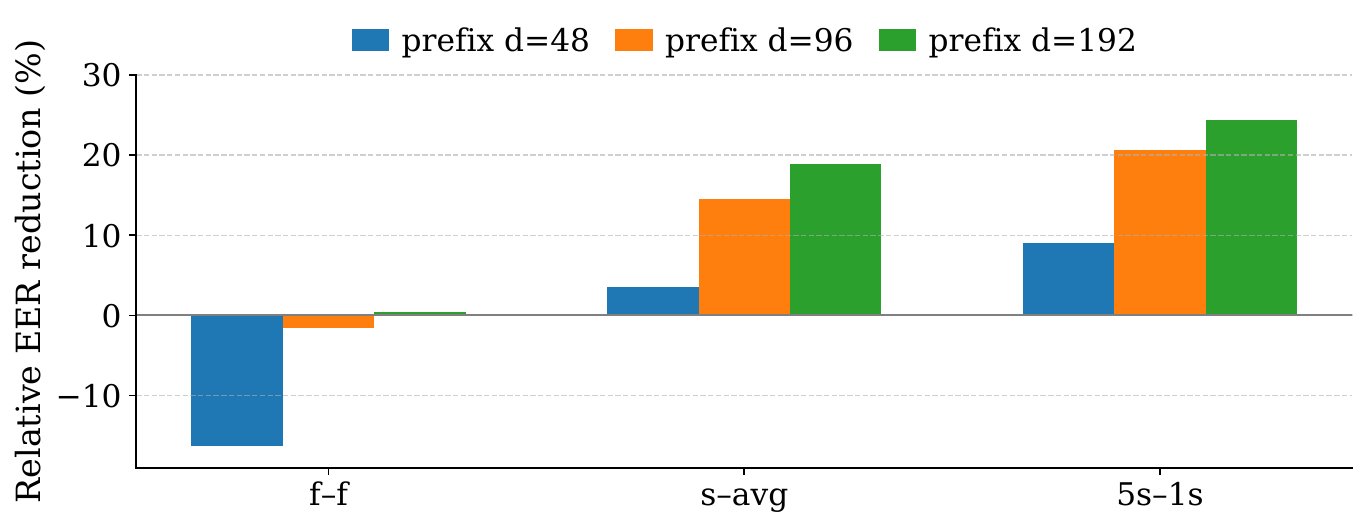}
  \vspace{-0.25cm}
  \caption{Relative EER reduction (\%) vs. ECAPA-TDNN baseline ($D{=}192$) on VoxCeleb1-O for one DAME-GT-SW model at prefixes $d\in\{48,96,192\}$; results are shown for \ff, \savg, and 5s\text{--}1s.}
  \label{fig:vox1o_rel_impr_vs_baseline}
  \vspace{-0.25cm}
\end{figure}


\subsection{Prefix analysis}
Figure~\ref{fig:vox1o_rel_impr_vs_baseline} shows the relative EER reduction of the DAME-GT-SW model compared to the ECAPA-TDNN baseline across prefixes $d\in\{48,96,192\}$.
With the smallest prefix ($d=48$), the DAME-GT-SW was superior to the baseline on the short-duration trials (s--avg $+3.5\%$, 5s\text{--}1s $+9.1\%$), but was inferior on the f--f condition ($-16.3\%$).
However, as the prefix dimension increased ($d=96$ or $d=192$), the performance gap observed on the f--f condition was significantly reduced, and the short-duration gains were enlarged.
These trends suggest that (i) short-duration cues were concentrated in leading dimensions via the DAME’s intended capacity–duration alignment, and (ii) higher dimensions were crucial for long utterances, thus supporting the use of a full embedding at test time.

\subsection{Domain robustness on VOiCES}
Table~\ref{tab:voices} summarizes the results on the VOiCES dataset, which contains far-field speech recorded in noisy and reverberant room environments.
Given the severe domain mismatch between training and test, these results assess the systems' domain robustness.
In the general training (GT) regime, the VLT generally improved the baseline on both s--avg and 5s--1s conditions, but with substantial EER increases on the f--f condition.
Similarly, the DAME-GT-SW degraded the baseline on the f--f condition, but with much smaller EER increases compared to the VLT with ECAPA-TDNN and ERes2NetV2 models.
In contrast, the DAME-GT-SW consistently outperformed all other systems on both s--avg and 5s--1s conditions across all encoder architectures.

In the fine-tuning (FT) regime, all methods improved the baseline on the f--f condition, except for the DAME-FT-SW, which used only short ($\leq 2$ s) utterances during fine-tuning.
Similar to that observed in Table~\ref{tab:ft-compact-split}, the LMFT still degraded the SV performance on the short-duration trials.
In contrast, the DAME-FT-HW generally improved the baseline on both s--avg and 5s--1s conditions across various encoder architectures, while consistently outperforming the D-ALMFT method (except for the 5s--1s condition with the ResNet34).
This suggests that the proposed duration-aligned prefix dimension supervision strategy improves generalizability over varying training utterance durations under domain mismatch, unlike the conventional D-ALMFT that fits the entire embedding dimensions toward both short- and long-duration utterances during the fine-tuning procedure.

To summarize, the results on VOiCES demonstrate that DAME's benefits are not limited to the in-domain VoxCeleb data.
Both DAME-GT-SW for general training and DAME-FT-HW for fine-tuning served as reliable strategies for enhancing short-duration SV robustness under significant domain mismatch, all without altering the speaker encoder architecture or increasing inference cost.




\newcommand{\mrowc}[2]{\multirow{#1}{*}[-1.1ex]{#2}}
\newcommand{\Mrowc}[2]{\multirow{#1}{*}[-5.9ex]{#2}}

\begin{table}[t]
\centering
\begingroup
\setlength{\abovecaptionskip}{6pt}   
\setlength{\belowcaptionskip}{6pt}   
\setlength{\textfloatsep}{12pt plus 2pt minus 2pt} 
\setlength{\floatsep}{10pt plus 2pt minus 2pt}     
\renewcommand{\arraystretch}{0.58}   
\setlength{\tabcolsep}{3pt}          
\vspace{-0.2cm}
\caption{VOiCES under domain mismatch (EER \%).}
\label{tab:voices}
\vspace{+0.1cm}
\resizebox{0.92\columnwidth}{!}{
\begin{tabular}{l l l ccc}
\toprule
Regime & Encoder & Method & \ff & \savg & 5s--1s \\
\midrule
\Mrowc{15}{GT}
& \mrowc{5}{ResNet34}
  & Baseline         & \textbf{5.30} & 11.02 & 17.50 \\
& & +VLT             & 5.78 & 10.59 & 15.63 \\
& & MRL              & 5.52 & 10.87 & 16.91 \\
& & {DAME-GT-SW}     & 5.75 & \textbf{10.55} & \textbf{15.60} \\
& & DAME-GT-HW       & 5.98 & 11.37 & 16.31 \\
\cmidrule(lr){2-6}
& \mrowc{5}{ECAPA-TDNN}
  & Baseline         & \textbf{4.84} & 11.18 & 18.53 \\
& & +VLT             & 5.96 & 11.25 & 17.09 \\
& & MRL              & 5.23 & 11.40 & 18.33 \\
& & {DAME-GT-SW}     & 5.24 & \textbf{10.79} & \textbf{16.34} \\
& & DAME-GT-HW       & 5.67 & 11.09 & 17.33 \\
\cmidrule(lr){2-6}
& \mrowc{5}{ERes2NetV2}
  & Baseline         & 5.21 & 10.99 & 17.09 \\
& & +VLT             & 5.76 & 10.81 & 16.09 \\
& & MRL              & \textbf{5.02} & 10.61 & 16.62 \\
& & {DAME-GT-SW}     & 5.31 & \textbf{10.25} & \textbf{15.53} \\
& & DAME-GT-HW       & 5.44 & 10.53 & 15.91 \\
\midrule
\Mrowc{15}{FT}
& \mrowc{5}{ResNet34}
  & Pretrained       & 5.30 & 11.02 & 17.50 \\
& & LMFT             & 4.97 & 11.49 & 18.63 \\
& & D-ALMFT          & 5.02 & 11.00 & 17.58 \\
& & DAME-FT-SW       & 6.11 & 11.01 & \textbf{16.43} \\
& & {DAME-FT-HW}     & \textbf{4.21} & \textbf{10.90} & 17.76 \\
\cmidrule(lr){2-6}
& \mrowc{5}{ECAPA-TDNN}
  & Pretrained       & 4.84 & 11.18 & 18.53 \\
& & LMFT             & \textbf{3.93} & 13.46 & 22.67 \\
& & D-ALMFT          & 4.26 & 11.16 & 18.61 \\
& & DAME-FT-SW       & 5.10 & 10.82 & 17.40 \\
& & DAME-FT-HW       & 4.18 & \textbf{10.61} & \textbf{17.12} \\
\cmidrule(lr){2-6}
& \mrowc{5}{ERes2NetV2}
  & Pretrained       & 5.21 & 10.99 & 17.09 \\
& & LMFT             & 4.96 & 11.22 & 17.88 \\
& & D-ALMFT          & 5.01 & 10.85 & 17.04 \\
& & DAME-FT-SW       & 5.63 & 11.39 & 17.21 \\
& & DAME-FT-HW       & \textbf{4.30} & \textbf{10.61} & \textbf{16.83} \\
\bottomrule
\end{tabular}}
\endgroup
\vspace{-0.3cm}
\end{table}

\section{Conclusions}
We introduced DAME, a model-agnostic framework that aligns embedding capacity with utterance length by training nested prefix embeddings with duration-supervised losses. Operating at the embedding level and orthogonal to MFA, DAME was applied via two efficient schemes, soft-weighted general training (DAME-GT-SW) and hard-weighted fine-tuning (DAME-FT-HW), without increasing inference cost. Experiments across three encoders on VoxCeleb1-O/E/H and the mismatched VOiCES dataset consistently showed improved short-duration EER while preserving full-length performance. Notably, DAME-GT-SW surpassed standard MRL variants, and DAME-FT-HW served as a superior alternative to LMFT for short-duration trials. Future work includes adaptive test-time prefix selection and joint duration–margin scheduling.

\clearpage
\bibliographystyle{IEEEbib}
\bibliography{strings,refs}

\end{document}